\documentclass[a4paper]{article}

\usepackage[english]{babel}
\usepackage[utf8x]{inputenc}
\usepackage[T1]{fontenc}

\usepackage[a4paper,top=3cm,bottom=2cm,left=3cm,right=3cm,marginparwidth=1.75cm]{geometry}

\usepackage{setspace}
\usepackage{amssymb}

\usepackage{cite}

\usepackage{amsmath}
\usepackage{graphicx}
\usepackage[colorinlistoftodos]{todonotes}

\title{Fourier single-pixel imaging in the terahertz regime}

\author{Rongbin She$^{1,2,\dagger}$, Wenquan Liu$^{1,\dagger}$, Yuanfu Lu$^{1,*}$, Zhisheng Zhou$^1$, and Guangyuan Li$^{1,*}$}

\date{}
\begin{document}
\maketitle

\begin{spacing}{2.0}

\noindent \large$^1$Shenzhen Institutes of Advanced Technology, Chinese Academy of Sciences, Shenzhen 518055, Guangdong Province, China

\noindent  $^2$Shenzhen College of Advanced Technology, University of Chinese Academy of Sciences, Shenzhen 518055, Guangdong Province, China

\noindent $^\dagger$ These authors contributed equally.

\noindent *yf.lu@siat.ac.cn; gy.li@siat.ac.cn

\end{spacing}

\begin{abstract}
We demonstrate Fourier single-pixel imaging in the terahertz regime. The experimental system is implemented with a photo-induced coded aperture setup, where a monolayer graphene on a high-resistance silicon substrate illuminated by a coded laser beam works as a terahertz modulator. Results show that high-quality terahertz images can be reconstructed using greatly reduced number of measurements. We further find that deep photo-induced terahertz modulation by adding a monolayer graphene on the silicon substrate and by using high laser power can significantly improve the image quality. Compared to Hadamard single-pixel imaging with re-ordered Hadamard matrix, the Fourier approach has higher image quality. We expect that this work will speed up the efficiency of single-pixel terahertz imaging and advance terahertz imaging applications.
\end{abstract}

Terahertz wave refers to electromagnetic radiation with a frequency of 0.1 THz -- 10 THz (corresponding to wavelength of 3 mm -- 30 $\mu$m). Terahertz imaging is widely used in security \cite{Shen2005THzImagingExplosive,Skvortsov2014THzImagingExplosive}, industrial inspection \cite{Duling2009THzImagingHidden,Koch2011THzReview}, and material composition identification \cite{XCZhang2006THzImagingExplosive,Shen2011THzPharmReview} because of its unique properties such as superior spatial resolution, perspective, and spectroscopic fingerprints \cite{Kawase2003THzImagingFinger,Chan2007THzImagingReview,Tonouchi2007THzReview}. One challenge in terahertz imaging applications has been the amount of time it takes to form an image \cite{Mittleman2018THzImagingReview}. In early terahertz imaging systems, image formation relied upon a pixel-by-pixel acquisition using a single-pixel terahertz detector and mechanical scanning \cite{Mittleman2018THzImagingReview}. However, this process is usually slow, limiting the image acquisition speed \cite{Mittleman2018THzImagingReview}. Therefore, efforts have been put on focal-plane array detection based on multi-pixel detectors in integrated form \cite{Ojefors2009THzImagingArray}. Despite of great progress, commercially available focal plane array detectors are still too expensive to afford for most researchers. As a result, approaches to speed up image acquisition with single-pixel terahertz detectors, which overcome the complexities and have a superior detection performance than
a multiple-pixel detector \cite{Abbott2014THzImagingArray}, have also been attracting increasing attention over the years.

In 2008, Chan {\sl et al.} \cite{Chan2008THzCS} proposed and demonstrated single-pixel terahertz imaging based on a powerful approach known as compressed sensing (CS): only 500 measurements are required in order to reassemble the full image of 4096 pixels. This approach replaces the mechanical scanning with the spatial terahertz modulation and thus enables a much shorter image acquisition time. Chan {\sl et al.} \cite{Chan2008THzCS2} further improved this approach by using a set of binary metal masks to modulate the object information, and reconstructed the full image of 1024 pixels with 300 measurements. In order to avoid the use of moving masks, which are difficult to align, Watts {\sl et al.} \cite{Watts2014THzCSMM} made use of metamaterial spatial modulator, Sensale-Rodriguez {\sl et al.} \cite{Xing2013grapheneTHzImaging} employed arrays of graphene electro-absorption modulators, and Kannegulla {\sl et al.} \cite{Kannegulla2014THzCAI} and Shams {\sl et al.} \cite{Shams2014THzCS} demonstrated photo-induced coded-aperture imaging using programmable illumination from a commercial available digital light processing projector. Recently, near-field terahertz imaging with subwavelength resolution \cite{Stantchev2016THzNearFieldImaging} were also demonstrated. Although great progress has been achieved, CS-based single-pixel terahertz imaging still suffers from long reconstruction time especially for large images\cite{Assmann2013CSImaging}.

Recently, Zhang {\sl et al.} \cite{Zhang2009SFSoptical} proposed and demonstrated Fourier single-pixel imaging (FSI) in the visible regime. They further showed that this approach can achieve high quality image with reduced reconstruction time \cite{Zhang2017Fast}. However, in the terahertz regime, this efficient approach has not been demonstrated yet.

Here, we experimentally demonstrate FSI in the terahertz regime, and clarify the differences compared with the FSI in the visible regime. We will implement a terahertz FSI system with a photo-induced coded aperture setup, where a monolayer graphene on a high-resistance silicon (GOS) substrate illuminated by a continuous wave (CW) laser works as a terahertz modulator. By investigating the reconstructed image quality under different sampling ratios, we will show that the FSI can greatly reduce the number of measurements while maintaining a high image quality. The effects of the photo-induced terahertz modulation depths on the image quality will also be discussed by comparing different laser powers that are delivered onto the GOS, and by comparing the GOS and the conventional high-resistance silicon (HRS) substrates. We will also compare the performance of terahertz FSI and Hadamard single-pixel imaging (HSI), the latter of which makes use of re-ordered Hadamard matrix.

\begin{figure}[htbp]
\centering\includegraphics[width=\linewidth]{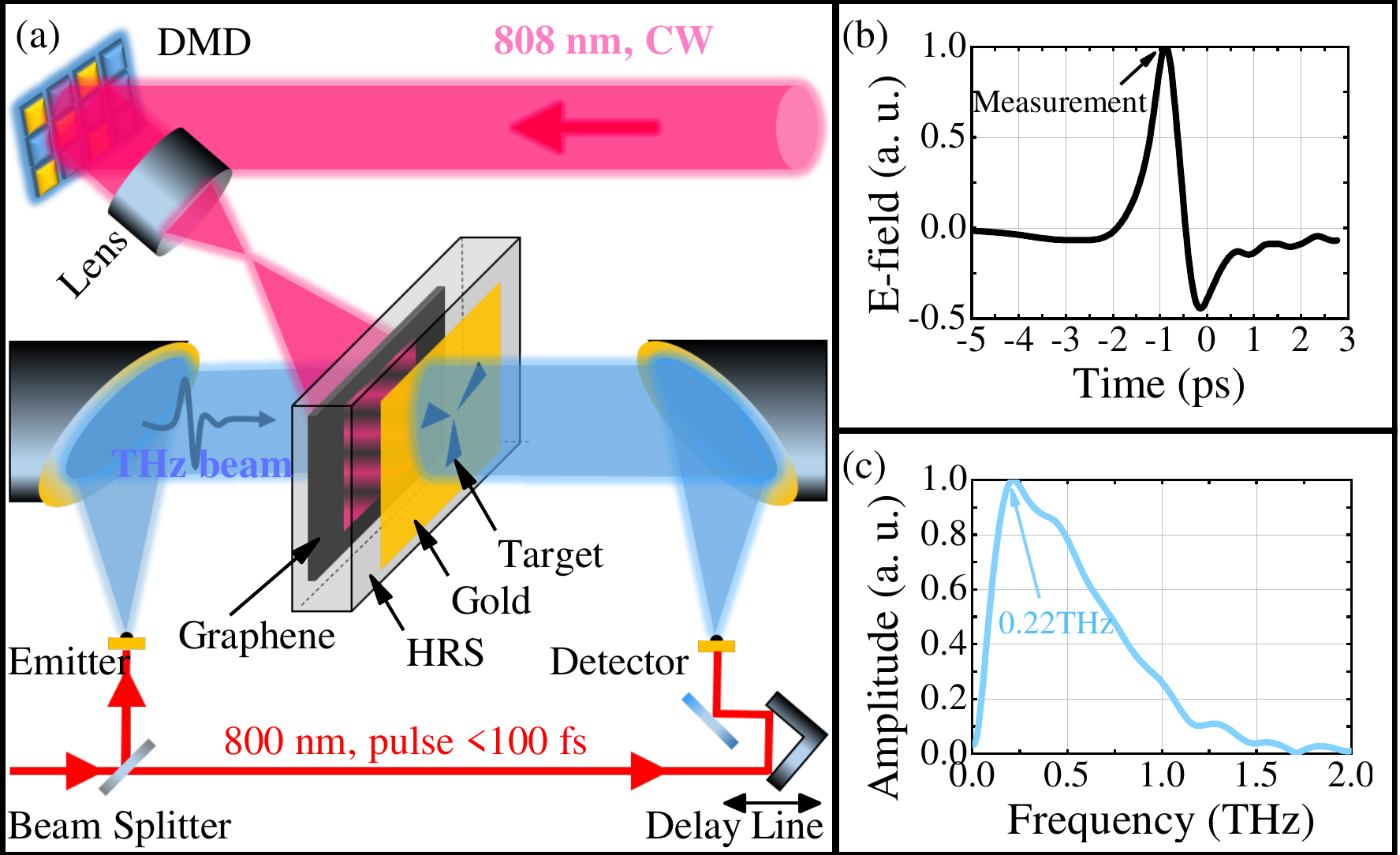}
\caption{(a) Schematic of the experiment setup for the terahertz FSI system. The target object is a three-arm cartwheel hollow-carved patterned in the gold film. (b) E-field and (c) normalized Fourier transform of the generated terahertz radiations.}
\label{fig:Setup}
\end{figure}

According to the theory of FSI in the visible regime, which can be found in refs  \citenum{Zhang2009SFSoptical,Zhang2017Fast,Zhang2017Hadamard}, we experimentally implemented a terahertz FSI system based on a typical home-built terahertz time-domain spectroscopy (THz-TDS) setup \cite{Liu2016THzTDS}. As illustrated in Fig.~\ref{fig:Setup}(a), a pair of biased low-temperature-grown GaAs photoconductive antennas illuminated by femtosecond laser pulses ($\lambda=800$ nm, $< 100$ fs, 80 MHz; Coherent Vitesse 800-5) are used to generate and detect terahertz radiations. Figure~\ref{fig:Setup}(b)(c) show the THz pulse and its spectrum. We recorded the measurements at the peak of the terahertz field, which was achieved by properly tuning the delay-line for femtosecond laser pulses. The central frequency of our terahertz pulse is $\sim$0.22 THz, corresponding to a central wavelength of 1.363 mm. The terahertz beam with a diameter of 3.82 mm is delivered onto a GOS substrate. The target object is a layer of 100 nm thick gold film with a three-arm cartwheel pattern, which was fabricated on the other side of the substrate through photolithography and thermal evaporation.

The GOS was prepared by transferring a monolayer graphene, which was grown on a copper substrate using chemical vapor deposition, onto a piece of HRS substrate (2000 $\Omega \cdot $cm, 100 $\mu$m thick) using polymethyl methacrylate and wet etching. Besides the GOS, we also had another piece of HRS working as the reference substrate. Note that these two pieces of HRS substrates were cut from the same wafer.

For the visible FSI system\cite{Zhang2009SFSoptical}, the incident light is directly coded with a digital micromirror device (DMD). For the terahertz FSI system, however, a photo-induced coded aperture setup is required, which can be implemented by projecting a CW 808 nm laser beam (of beam size $4 \,{\rm mm} \times 4 \,{\rm mm}$) encoded by a DMD (unmounted from a DLP3000, Texas Instruments) onto the GOS substrate. The photo-induced free charge carriers in the GOS leads to a change of the electric conductivity, which is inherently correlated with a decrease of the THz transmission
through the material \cite{Weis2012GOSmodulator}. Since the FSI sampling procedure requires masks with greyscale values within $[0, 1]$, here we used the video mode of the DMD, which can project greyscale values via temporal dithering at the cost of sacrificing the mask switch rates. The DMD can display 8-bit (256 quantization levels) grayscale patterns: 100\% intensity denotes white (255 grayscale), and 0\% denotes black (0 grayscale). Since a bright or dark laser spot encoded by the DMD induces low or high terahertz transmission through the GOS substrate, the grayscale encoded mask in the laser beam corresponds to a terahertz pattern with flipped grayscale values.

In other words, although the visible FSI and the terahertz FSI share the same theory, the experimental implementations are quite different: for the former, the incident light encoded by the projection masks can be directly used for imaging; whereas for the latter, the terahertz wave for imaging requires photo-induced modulation by an encoded laser beam.

For a terahertz image of $N\times N$ pixels, the FSI undersampling process is performed through $M$ measurements with $M\ll N$. The sampling ratio $S_{\rm R}$ can be defined as $S_{\rm R} \equiv M/N^2$. Thus a full sampling ratio means $M=N^2$.

In this work, we focus on the acquisition of $64 \times 64 $ terahertz images of the three-arm cartwheel target, {\sl i.e.}, $N=64$. Unless otherwise specified, the measurements were performed with the GOS substrate that is illuminated by a high laser power of 350 mW. The effects of the modulation depth will be discussed
later by imposing different laser powers and comparing the
GOS and HRS substrates.

\begin{figure}[htbp]
\centering\includegraphics[width=\linewidth]{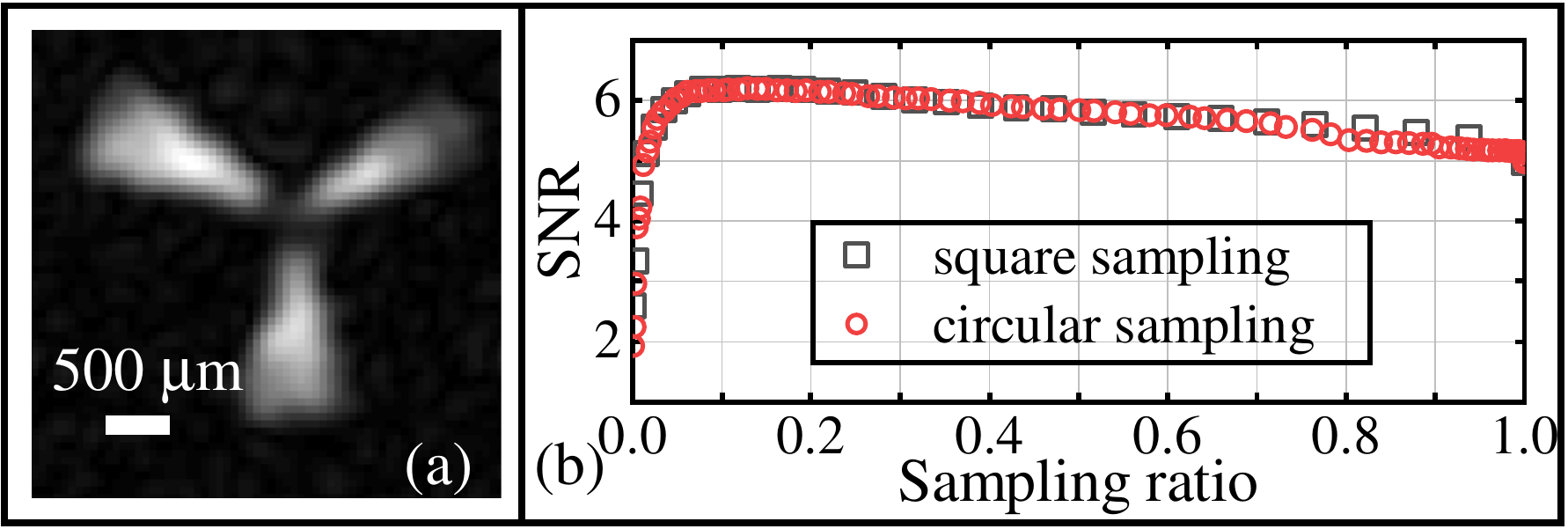}
\caption{(a) Image reconstructed from the terahertz FSI system at a sampling ratio of $S_{\rm R}=11.8\%$. (b) Signal-to-noise ratio (SNR) {\sl versus} sampling ratio for the square (symbols $\Box$) and circular (symbols $\circ$) undersampling schemes.}
\label{fig:SNR}
\end{figure}

Figure \ref{fig:SNR}(a) shows a $64\times 64$ image obtained from the terahertz FSI system at a sampling ratio of $S_{\rm R}=11.8\%$. Results show that the outline of the three-arm cartwheel, except for the central regions with small width, can be reconstructed. The resolution is limited by our imaging setup, and is mainly limited by the thickness of our GOS photomodulator (100 $\mu$m). In order to calculate the diffraction limit of our  experimental system, we follow the method outlined in ref \citenum{Stantchev2016THzNearFieldImaging} and elaborated in ref \citenum{Kowarz1995Diff}. For our experimental setup with central wavelength of 1.363 mm and 100 $\mu$m thick GOS substrate, calculations show that the resolution is about 270 $\mu$m. Therefore, structures smaller than 270 $\mu$m cannot be distinguished, consistent with our experimental results. In other words, the experimental setup also acts as a low-pass spatial filter.

In order to quantify the image quality, we adopt signal-to-noise ratio (SNR) as the figure of merit following refs \citenum{Stantche2017THzNearFieldImaging} and \citenum{Shchelokova2018THzImagingSNR}. It is defined as the average value of the signal in the region of the three-arm cartwheel, $\mu({\rm signal})$, divided by the standard deviation of the noise in the signal-free area of the image, $\sigma({\rm background})$, that is\cite{Stantche2017THzNearFieldImaging,Shchelokova2018THzImagingSNR}
\begin{equation}
\label{eq:SNR}
{\rm SNR} \equiv \frac{\mu({\rm signal})}{\sigma({\rm background})}.
\end{equation}

\begin{figure}[htbp]
\centering\includegraphics[width=\linewidth]{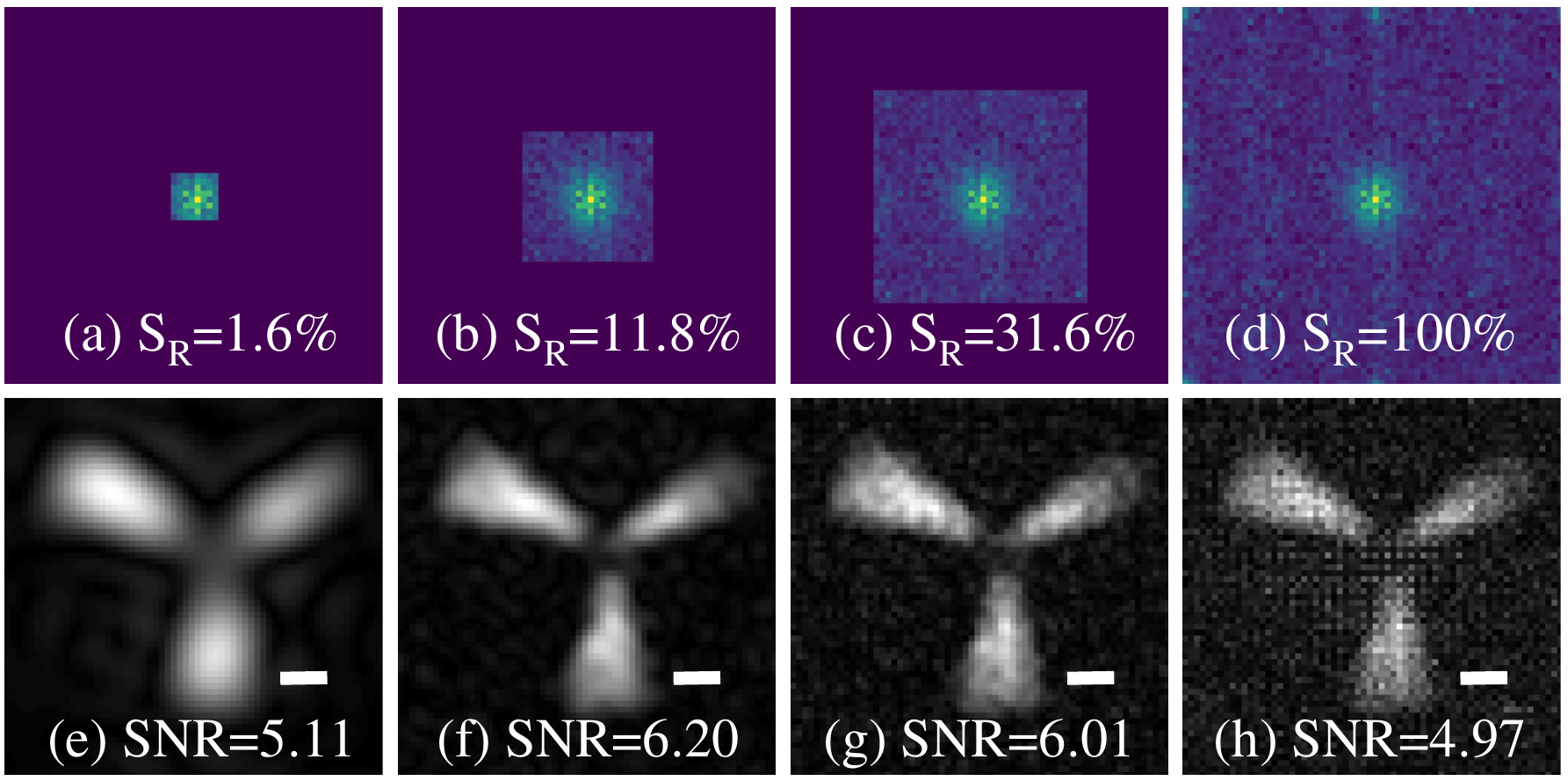}
\caption{(a)--(d) Truncated (based on square undersampling scheme) spatial Fourier spectrum of different sampling ratios in terahertz FSI experiments. (e)-(h) The corresponding reconstructed images with SNRs labeled. Each column has the same sampling ratio, and scalar bars in (e)--(h) denote 500 $\mu$m.}
\label{fig:Square}
\end{figure}

Figure \ref{fig:SNR}(b) shows that the SNR of the reconstructed images first quickly increases and then slowly decreases as the sampling ratio increases. The SNR results show that the image can be reconstructed even for a very small sampling ratio down to 1.6\%. The corresponding sampling of the spatial Fourier spectrum and the reconstructed image are shown by Figs.~\ref{fig:Square}(a) and \ref{fig:Square}(e), respectively. In this scenario, the reconstructed image is very bright compared with the clean and dark background, corresponding to a large SNR of 5.11. A problem is that the edge of the reconstructed image is not very clear. This is because the low-spatial-frequency information only contains the large area variations, which determine the global image contrast. In other words, the target object's global outline can be accurately reconstructed even for a very small sampling ratio. Note that the undersampling process for the FSI system is performed by truncating the spatial Fourier spectrum with a low-pass filter, as illustrated in Figs.~\ref{fig:Square}(a)--(d). As the sampling ratio increases to $S_{\rm R} = 4.8\%$, Fig.~\ref{fig:SNR}(b) shows that the SNR dramatically increases to 6.00. The SNR further slowly increases to the maximum value of 6.20 at $S_{\rm R} =11.8\%$. Meanwhile, the edges of the reconstructed images become clearer, as shown by Fig.~\ref{fig:Square}(f). This is because an increased sampling ratio includes more high-spatial-frequency information, which contains the small area variations and determines the sharpness and local image contrast. However, as the sampling ratio further increases, the SNR gradually decreases to 4.97 at full sampling, because periodic noises of high spatial frequency are included for larger sampling ratios\cite{Zhang2009SFSoptical}, as shown in Fig.~\ref{fig:Square}(g)--(h). Remarkably, the SNR of the Fourier system is larger than 6.0 for $4.8\% \leq S_{\rm R} \leq 31.6\%$. These striking performances, {\sl i.e.}, high image quality at low sampling ratio, indicate that the FSI approach is very efficient in the terahertz regime.

\begin{figure}[htbp]
\centering\includegraphics[width=\linewidth]{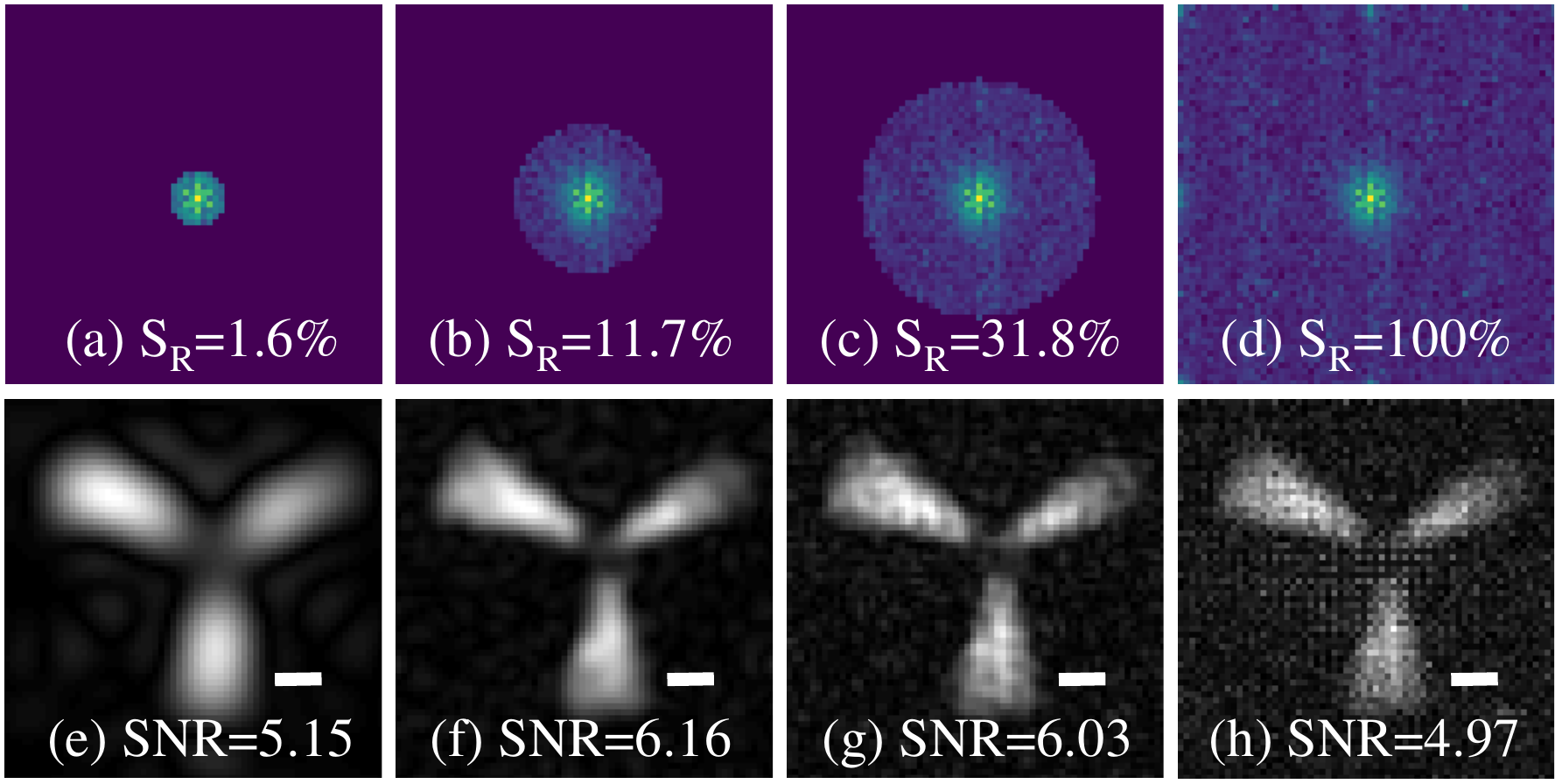}
\caption{Similar to Fig.~\ref{fig:Square} but for the circular undersampling scheme.}
\label{fig:circular}
\end{figure}

In the above discussion, we have adopted square undersampling scheme of the spatial frequency spectrum. One may speculate that the slightly decreasing SNR for large $S_{\rm R}$ would originate from the sharp edges of the square undersampling region. In order to clarify this issue, we also performed circular undersampling scheme, as shown in Fig.~\ref{fig:circular}. Fig.~\ref{fig:SNR}(b) shows that the circular and square undersampling schemes have almost the same SNR behaviors, as further validated by comparing Figs.~\ref{fig:Square} and \ref{fig:circular}. Therefore, the sharp edges of the square undersampling region do not decrease the quality of the reconstructed images.

We note that the optimal sampling ratio for the maximum SNR may change if different target objects are imaged. For example, for a six-arm cartwheel (not shown here due to the space limitation), we found that the optimal sampling ratio for the maximum SNR is now 9.8\% for the square undersampling scheme or 9.2\% for the circular undersampling scheme. We emphasize that these extremely low sampling ratio for high quality image is an intrinsic characteristic for the  terahertz FSI approach. This is because our experimental system with a resolution of 270 $\mu$m acts as a low-pass spatial filter and the spatial frequency spectrum of the image is further truncated with a low-pass filter, whereas the standard deviation of the white noises, which have high-spatial frequencies, will be small for low sampling ratios.

We now study the effects of the photo-induced terahertz modulation depth by varying the laser power impinging onto the GOS substrate, and by comparing the GOS and the HRS substrates under the same laser power. For these comparisons, the square undersampling scheme with $S_{\rm R}=11.8\%$ is used.

\begin{figure}[htbp]
\centering\includegraphics[width=\linewidth]{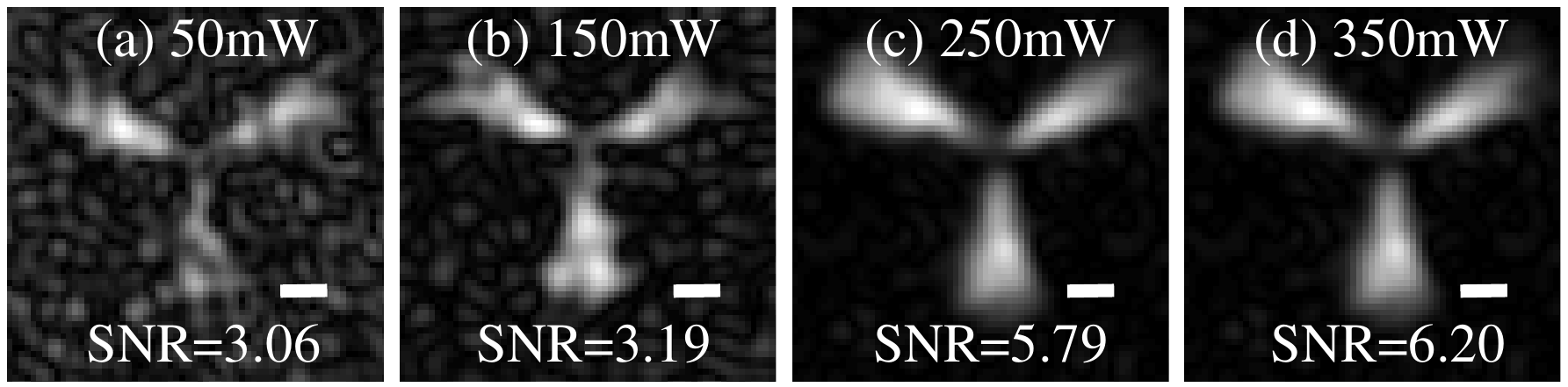}
\caption{Images reconstructed from the terahertz FSI system using the GOS substrate at different laser powers: (a) 50 mW, (b) 150 mW, (c) 250 mW, and (d) 350 mW. Scalar bars denote 500 $\mu$m.}
\label{fig:Power}
\end{figure}

Figure \ref{fig:Power} shows that the SNR of the reconstructed terahertz images increases with the laser power: from ${\rm SNR}=3.06$ for 50 mW increases 3.19 for 150 mW, 5.79 for 250 mW and 6.20 for 350 mW. This is because as the laser power increases, higher carrier density is generated in the GOS, resulting in deeper terahertz modulation. On the other hand, a larger contrast between “1” (high terahertz transmission) and “0” (low terahertz transmission) will enhance the spatial frequency information of the target object compared with the noises.

\begin{figure}[htbp]
\centering\includegraphics[width=0.5\linewidth]{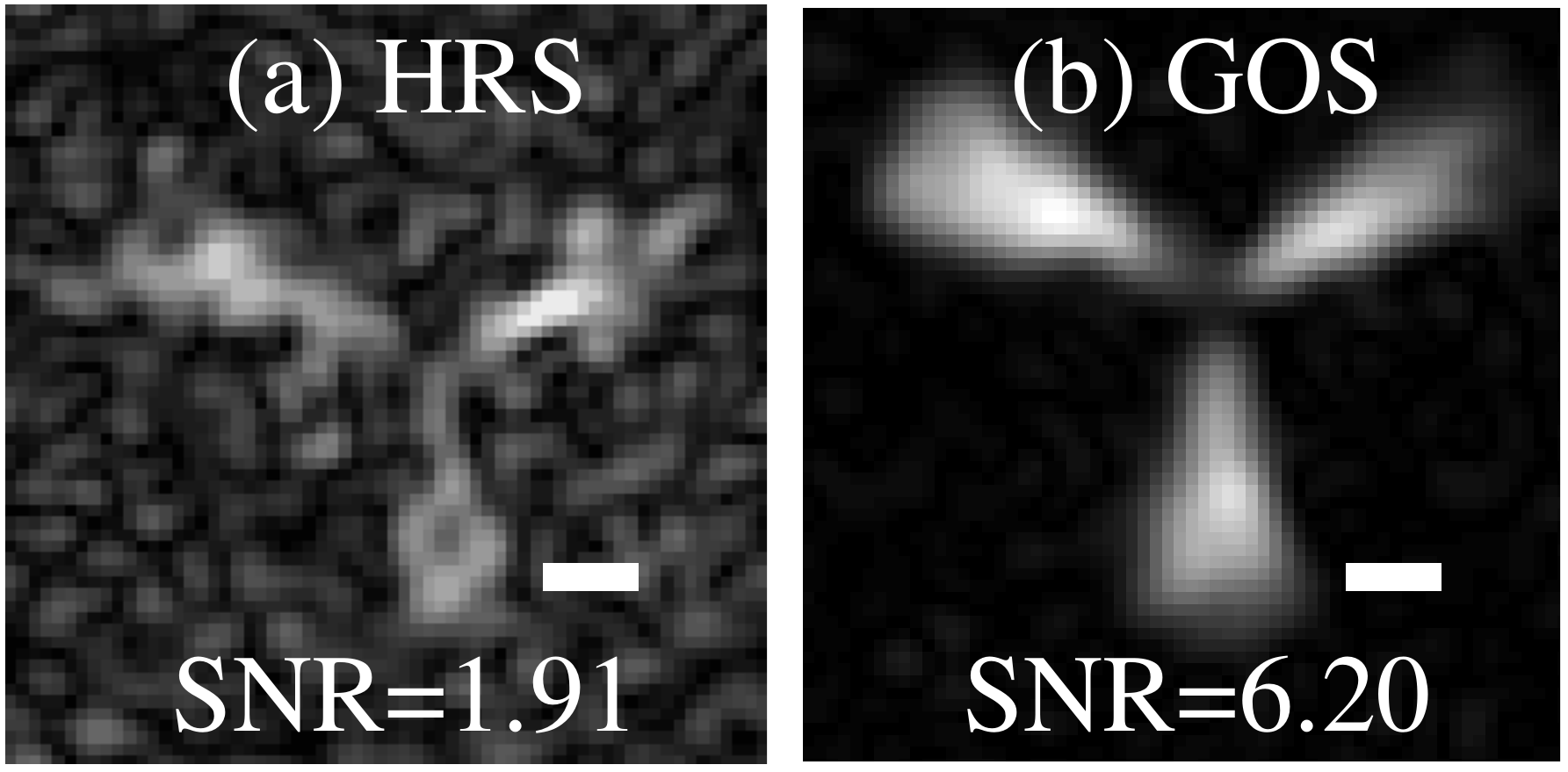}
\caption{Similar to Fig.~\ref{fig:Power} but for (a) HRS and (b) GOS substrates at the laser power of 350 mW.}
\label{fig:Sub}
\end{figure}

Figure \ref{fig:Sub} compares the reconstructed terahertz images using the HRS and the GOS substrates under the same laser power of 350 mW. Results show that for the HRS substrate, the reconstructed image has very poor quality with a small SNR of only 1.91; whereas for the GOS substrate, the reconstructed image is very clean and clear, and the SNR is as high as $6.20$, more than three times of that for the HRS substrate. By comparing Figs.~\ref{fig:Power}(a) and \ref{fig:Sub}(a), we notice that the GOS substrate under a low laser power of 50 mW has much better image quality than the HRS substrate under a high laser power of 350 mW. This great improvement is due to the much deeper photo-induced terahertz modulation by using the GOS, as demonstrated in refs \citenum{Weis2012GOSmodulator,Liang2015grapheneTHzModul,Sun2018THzGraphene}.

\begin{figure}[htbp]
\centering\includegraphics[width=\linewidth]{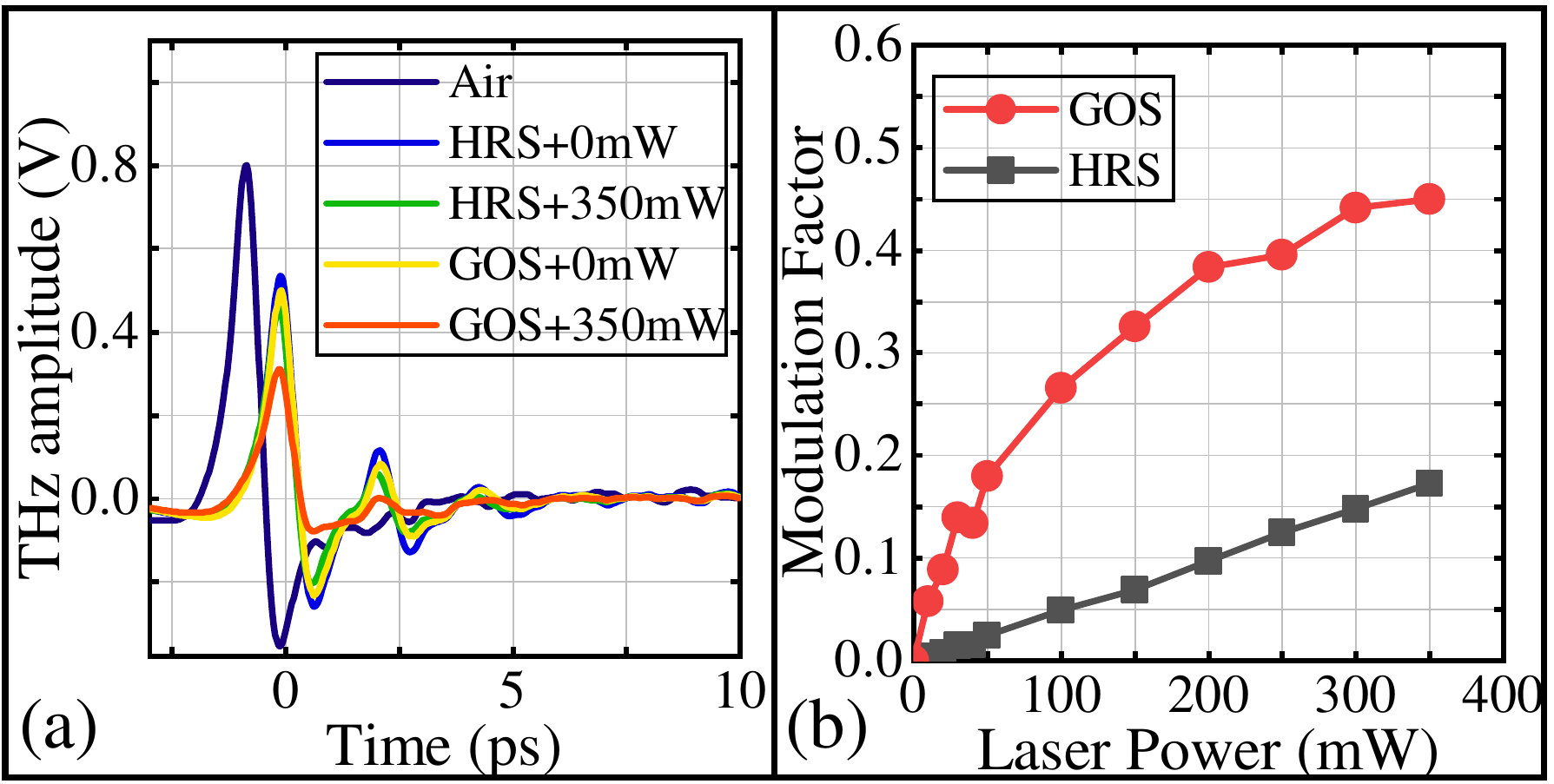}
\caption{(a) Time domain terahertz pulses and (b) modulation factors of the HRS and GOS substrates under different laser powers.}
\label{fig:MF}
\end{figure}

The different photo-induced terahertz modulation depths of the GOS and the HRS substrates are better depicted by Fig.~\ref{fig:MF}. Under the irradiation of laser, the terahertz transmission through the GOS substrate is decreased by a larger magnitude than that through the HRS substrate. We adopt the modulation factor defined in ref \citenum{Guocui2017ModulationFactor} to quantify the modulation depth. Fig.~\ref{fig:MF}(b) shows that the modulation factor generally increases with the laser power for both the GOS and the HRS substrates. For different laser powers, the GOS always has larger modulation factors than the HRS.

Since for an optically controlled THz modulator, longer carrier lifetime benefits to significant modulation as it helps reach higher non-equilibrium carrier concentration \cite{DongOptically2017}. We further calculate the carrier lifetime for the GOS and the HRS substrates. Following ref \citenum{TingModulator2016}, we first extracted the transient
 complex photoconductivity at the frequency of 0.22 THz from the
time-domain data of the transmitted THz pulses, and then qualitatively analyzed the carrier density using the simple Drude model. The carrier lifetime $\tau$ can then be obtained from the carrier density $n$ through $n=I_0(1-R)\tau/(2Ad\hbar\upsilon)$, where $I_0$ is the average power of the incident CW laser, $A$ is the laser beam area, $d$ is the penetration depth, $R$ is the optical reflectivity of the substrate, and $\hbar\upsilon$ is the photon energy \cite{Shrekenhamer2013THzimagingchange}.
For $I_0=350$ mW and $A=16\,{\rm mm}^2$, the calculated carrier density of the HRS is found to be $n=6.05 \times 10^{16} \, {\rm cm}^{-3}$, and the calculated carrier lifetime is $\tau=23.3\,\mu$s. We note that this carrier lifetime is comparable to the measured value ($\sim25\,\mu$s) reported in the literature \cite{SchulenburgElectropassivation2000,GaubasComparative2007}. Similarly, for the GOS, we find that the calculated carrier density is $n=3.59\times 10^{17}\, {\rm cm}^{-3}$, and the calculated carrier lifetime is $\tau=138.1\,\mu $s, which is about six times of that for the HRS substrate. This longer carrier lifetime originates from the fact that photo-induced free electrons flow from the silicon to the interface of the monolayer graphene and the silicon, leading to larger modulation depth of terahertz wave \cite{Weis2012GOSmodulator,MaixiaEfficient,MultilayerKaya}.

By using the same experimental setup for the terahertz FSI system as illustrated in Fig.~\ref{fig:Setup}(a), we can also realize a terahertz Hadamard single-pixel imaging (HSI) system and compare its performance with the terahertz FSI system. The only differences between these two systems lie in the encoded mask and the reconstruction: the FSI makes use of masks in form of sinusoidal stripes with grayscale values, whereas the HSI masks are two-dimensional (2D) Hadamard projection matrices with binary values; the reconstruction for the FSI is achieved through inverse Fourier transform, whereas for the HSI, the reconstruction is achieved through inverse Hadamard transform. Note that here we adopt the terahertz HSI approach as the reference since it was shown to outperform the other single-imaging approaches quite recently \cite{YounessMatrices2017,Mattew2019Principles}. Moreover, since the comparison between the FSI and the HSI has been done in the visible regime \cite{Zhang2017Hadamard}, it is also worthwhile to do so in the terahertz regime.

For fair comparison, the 2D Hadamard matrix is re-ordered such that the Hadamard spectrum has similar spatial distribution as the Fourier spectrum: the undersampling process removes the high-spatial frequencies of the target object \cite{MinRussian2017,Zhang2017Hadamard}.

\begin{figure}[htbp]
\centering\includegraphics[width=\linewidth]{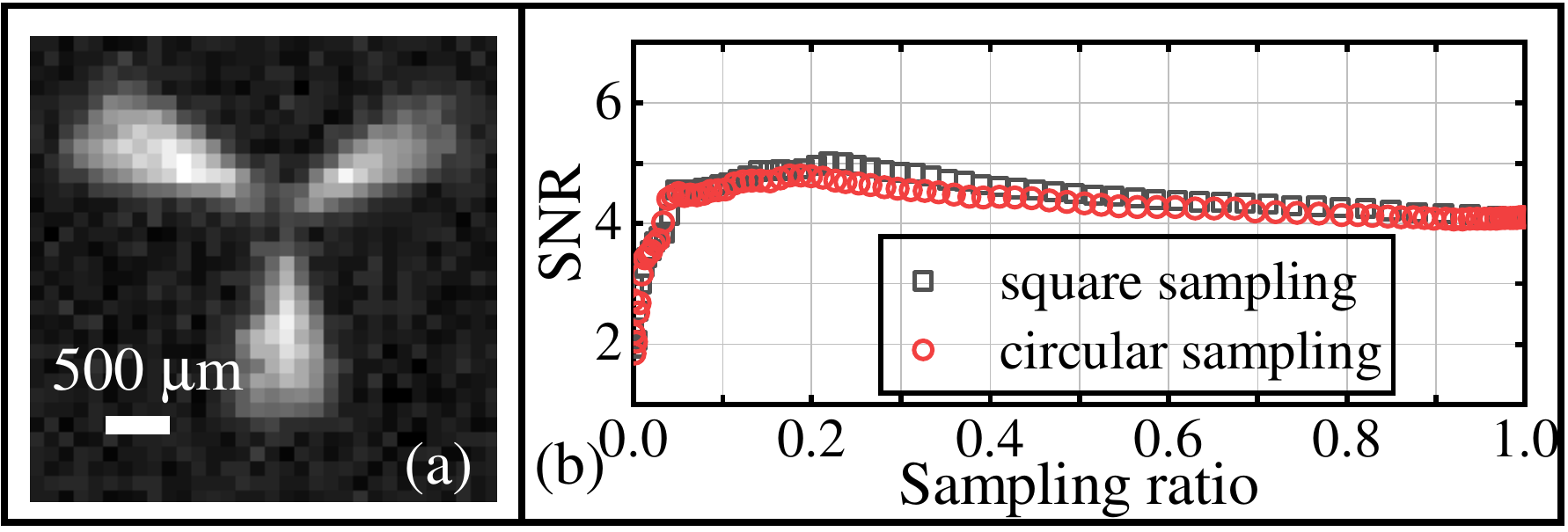}
\caption{(a) Image reconstructed from the terahertz HSI system with ${\rm SNR}=5.0$ at $S_{\rm R} =22\%$. (b) SNR {\sl versus} sampling ratio for the square (symbols $\Box$) and circular (symbols $\circ$) undersampling schemes.}
\label{fig:Hadamard}
\end{figure}

Figure~\ref{fig:Hadamard}(a) shows that, by using the HSI approach, a clear $64\times64$ terahertz image can be reconstructed with ${\rm SNR}=5.0$ at a sampling ratio of $S_{\rm R} =22\%$.
Fig.~\ref{fig:Hadamard}(b) shows that the SNR first increases to 5.0 at $S_{\rm R} =22\%$ and then slowly decreases to 4.07 at the full sampling ratio, and that both the square and circular undersampling schemes have almost the same performance. These behaviors are very similar to those for the terahertz FSI system, as shown in Fig.~\ref{fig:SNR}(b), except that the HSI system has smaller SNR values.

\begin{figure}[htbp]
\centering\includegraphics[width=\linewidth]{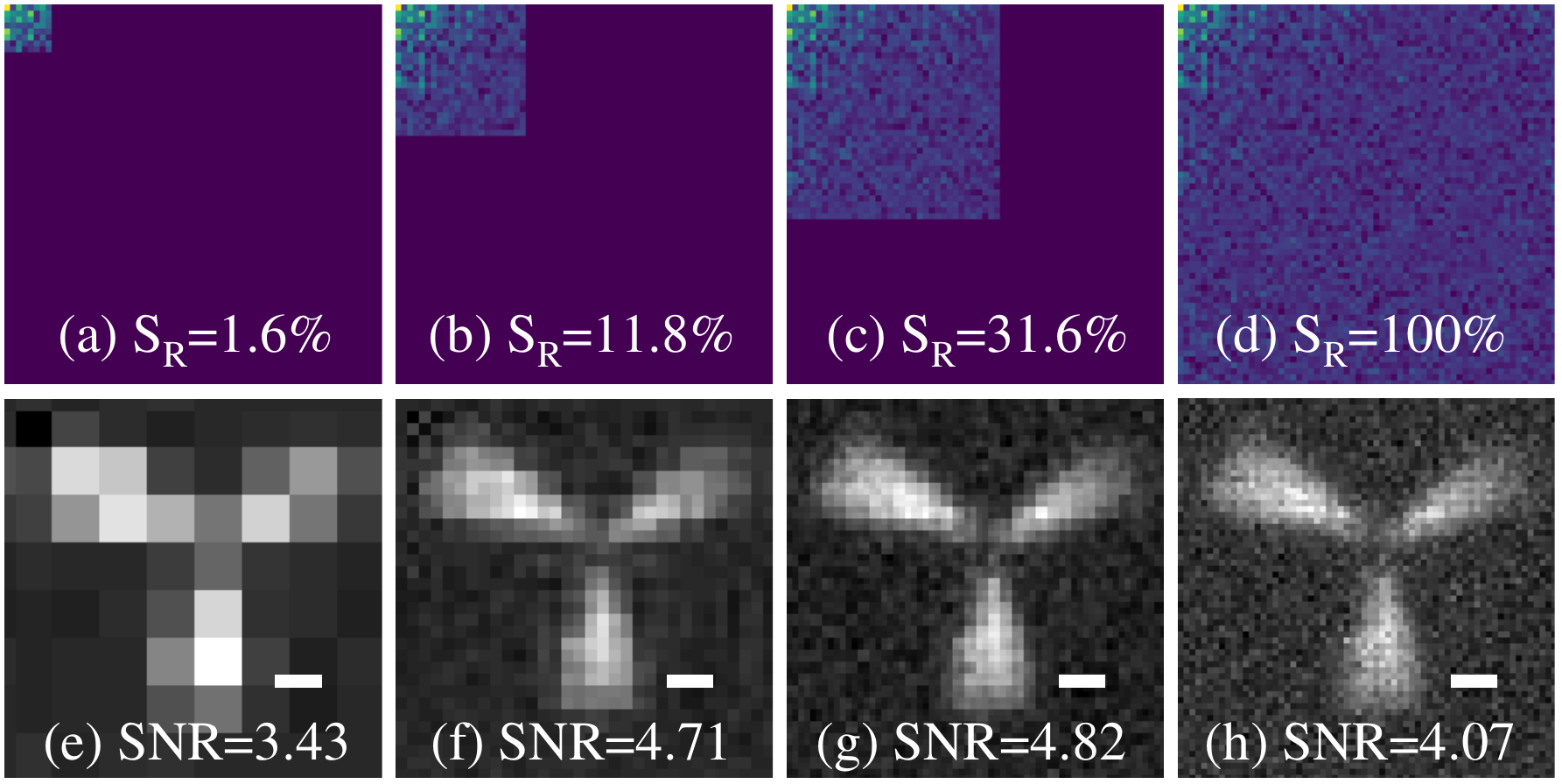}
\caption{Similar to Fig.~\ref{fig:Square} but for the terahertz HSI system.}
\label{fig:HadamardSquare}
\end{figure}

Indeed, these similar behaviors can be understood similarly. Fig.~\ref{fig:HadamardSquare} shows the square undersampling scheme of the Hadamard spectrum and the corresponding reconstructed terahertz images. The circular undersampling scheme has almost the same performance and thus is not shown here due to the space limitation. Results show that at a very low sampling ratio of $S_{\rm R} =1.6\%$, strong mosaic effect can be observed in the reconstructed image. For $S_{\rm R} =11.8\%$, the mosaic effect can still be observed, resulting in a signal-noise-ratio of ${\rm SNR}=4.71$, which is smaller than that of the counterpart for the terahertz FSI system: ${\rm SNR}=6.20$ (Fig.~\ref{fig:Square}(f)). As the sampling ratio further increases, the mosaic effect gradually disappear. Comparing Figs.~\ref{fig:Square} and \ref{fig:HadamardSquare}, we find that the terahertz FSI system can reconstruct images with higher quality under the same conditions, consistent with the results in the visible regime \cite{Zhang2017Hadamard}.

In conclusion, we have demonstrated FSI in the terahertz regime and clarified its differences from the approach in the visible regime. We have experimentally implemented a terahertz FSI system based on a photo-induced coded aperture setup, which can be built on a typical THz-TDS system. Results have shown that the reconstruction can be done with a large SNR of 5.11 even when the sampling ratio is as small as 1.6\%. More strikingly, the SNR can be larger than 6.0 when the sampling ratio is between 4.8\% and 31.6\%, and can reach a maximum of 6.20 at a sampling ratio of 11.8\%. We have found that the square and circular undersampling schemes have almost the same performance. By comparing different laser powers and comparing the GOS and the HRS substrates, we have shown that a large photo-induced terahertz modulation depth greatly improves the quality of the reconstructed images, pointing to strategies for further improvement. We have also performed comparison of FSI and HSI in the terahertz regime and found that the FSI performs better than the HSI. We expect our work will greatly improve the efficiency and the image quality of single-pixel terahertz imaging, and will advance terahertz imaging system in practical applications.

This work is supported by the Shenzhen Research Foundation (Grant Nos. JCYJ20150925163313898, JCYJ20160608153308846 and JCYJ20160510154531467), the National Key Research and Development Program of China (2017YFC0803506), and the Youth Innovation Promotion Association, CAS (No. 20160320).

\end{document}